\documentclass[a4paper,aps,prd,twocolumn,groupedaddress,nofootinbib,floatfix]{revtex4}
\usepackage{graphicx}
\usepackage{aas_macros}

\def\eq#1{(\ref{#1})}
\def\Tr{\mathrm{Tr}}
\newcommand{\G}{\mathcal{G}}
\newcommand{\p}{\mathcal{P}}
\newcommand{\D}{\mathcal{D}}

\newcommand{\expectWRT}[2]{\left\langle #1\right\rangle_{#2}}

\begin{document}
\title{Inference with minimal Gibbs free energy in information field theory}
\author{Torsten A. En{\ss}lin}
\author{Cornelius Weig}
\affiliation{Max-Planck-Institut f\"ur Astrophysik, Karl-Schwarzschild-Str. 1, 85741 Garching, Germany}
\date{\today}

\begin{abstract}
Non-linear and non-Gaussian signal inference problems are difficult to tackle. Renormalization techniques permit us to construct good estimators for the posterior signal mean within information field theory (IFT), but the approximations  and assumptions made are not very obvious. Here we introduce the simple concept of minimal Gibbs free energy to IFT, and show that previous renormalization results emerge naturally. They can be understood as being the Gaussian approximation to the full posterior probability, which has maximal cross information  with it. We derive optimized estimators for three applications, to illustrate the usage of the framework: (i) reconstruction of a log-normal signal from Poissonian data with background counts and point spread function, as it is needed for gamma ray astronomy and for cosmography using photometric galaxy redshifts, (ii) inference of a Gaussian signal with unknown spectrum and (iii) inference of a Poissonian log-normal signal with unknown spectrum, the combination of (i) and (ii). Finally we explain how Gaussian knowledge states constructed by the minimal Gibbs free energy principle at different temperatures can be combined into a more accurate surrogate of the non-Gaussian posterior.
\end{abstract}

\maketitle

\section{Introduction}
\label{sec:intro}
\subsection{Abstract inference problem}
\label{sec:generic inference problem}

Measurements provide information on the signals we are interested in, encoded in the delivered data. How can this information be best retrieved? Is there a generic and simple principle from which optimal data analysis strategies derive? Can an information energy be constructed which -- if minimized -- provides us with the correct knowledge state given the data and prior information? And if this exists, how can this information ground state be found at least approximatively?

An information energy, to be minimized, would be very useful to have, since many of the existing minimization techniques, analytical and numerical, can then be applied to it. A number of such functions to be extremized to solve inference problems were proposed in the literature, like the likelihood, the posterior, or the entropy. The likelihood is the probability that the data has resulted from some signal. The posterior is the reverse, it is the  probability that  given the data some signal was the origin of it. Extremizing either of them certainly makes sense, but often ignores the presence of slightly less probable, but much more numerous possibilities in the signal phase space. Those have a much larger entropy and are therefore favored by maximum entropy methods. However, maximum entropy alone can not be the inference determining criterion, since it favors states of complete lack of knowledge, irrespective of the data. Thus some counteracting energy is required which provides the right amount of force to the inference solution. Here, we argue that the ideal information energy is provided by the Gibbs free energy, which combines both maximum entropy and maximum a posteriori (MAP) principles. 

The Gibbs free energy has to be regarded as a functional over the space of possible probability density functions (PDF) of the signal given the data. The result of the minimization is therefore a PDF itself, and not a single signal estimate. Minimizing the Gibbs free energy maximizes the entropy within the constraints given by the internal energy. The latter is understood as the average of the negative logarithm of the joint probability function of signal and data weighted with the PDF.

The usage of thermodynamical concepts for inference problems is not new, see e.g. \cite{1957PhRv..106..620J, 1957PhRv..108..171J}. What is new here, is that we develop this for signals which are fields, spatially distributed quantities with an infinite number of degrees of freedom, while using an approximate Gaussian ansatz for the PDF to be inferred. We thereby connect information field theory (IFT) \cite{1985ApJ...289...10F, 1987ApJ...323L.103B, 1987PhRvL..58..741B, 1988PhRvL..61.1512B, 1996PhRvL..77.4693B, 1999physics..12005L, Lemm2003, 2009PhRvD..80j5005E}, as a statistical field theory dealing with a huge number of microscopic degrees of freedom, to thermodynamics, as a means to generate simplified, but macroscopic descriptions of our knowledge. 
Thereby we find that former IFT results obtained with complex renormalization schemes in \cite{2010arXiv1002.2928E,2009PhRvD..80j5005E} can easily be reproduced, and even be extended to more complicated measurement situations. 

In the remainder of Sect. \ref{sec:intro} we briefly introduce to IFT, MAP, and Maximum Entropy. This motivates the minimal Gibbs free energy principle, which we formally derive in Sect. \ref{sec:thermoinf}, and show its equivalence to maximal cross information. The application of this principle to optimize approximations of the posterior  of concrete inference problems is provided in Sect. \ref{sec:application}. There, the log-normal Poisson problem (Sect. \ref{sec:plndata}) and the problem to reconstruct without known signal power spectrum (Sect. \ref{sec:rec. without spec. knowledge}), as well as their combination (Sect. \ref{sec:LNPandspecrec}) are addressed. Finally, we show how approximate posteriors obtained at different temperatures can be combined into a better posterior surrogate in Sect. \ref{sec:infosynth} before we conclude Sect. \ref{sec:conclusions}.

\subsection{Information field theory}

Information theory describes knowledge states with probabilities. If $\Omega$ is the complete set of possibilities, and $A \subset \Omega$ is a subset, then $P(A)\in [0,1]$ describes the plausibility of $A$ being the case, with $P(A)=1$ denoting $A$ being assumed to be sure, $P(A)=0$  denoting $A$ being (assumed to be) impossible, and $0<P(A)<1$ describing uncertainty about the truth of $A$. Obviously $P(\Omega)=1$ and $P(\emptyset)=0$. The usual rules of probability theory apply, and generalize the binary logic of Aristotle to different degrees of certainty or uncertainty \cite{Cox1946,Cox1963}. In case the set of possibilities is a continuum, it makes sense to introduce a PDF $\p(\psi)$ over $\Omega$, so that  $P(A) = \int_A d\psi \, \p(\psi)$. Each possible state $\psi$ can be a multi-component vector, containing all aspects of reality which are in the focus of our inference problem. 

We might be interested in a sub-aspect of $\psi$ which we call our signal  $s=s(\psi)$.
The induced signal PDF is retrieved from a functional or path integral over all the phase spaces of the possibilities of $\psi$ via 
$P(s) = \int \D\psi \, \p(\psi)\, \delta(s-s(\psi))$.
If $s$ is a field, a function over a physical space $V$, then $s = (s_x)_{x\in V}$ might be a vector in the Hilbert space  $\Omega$ of all $L^2$-integrable functions over $V$ and $\p(s)$ is then a probability density functional. Information theory for $s$ becomes IFT, which is a statistical field theory. 

Inference on the signal $s$ from data $d$ is done from the posterior probability $\p(s|d)$, which can be constructed from the joint PDF of signal and data $\p(d,s)$ via
\begin{equation}\label{eq:psd}
 \p(s|d) = \frac{\p(d,s)}{\p(d)} = \left. \frac{e^{-\beta\,H(d,s)}}{Z_\beta}\right|_{\beta = 1},
\end{equation}
where $\p(d,s)=\int_\Omega \D\psi\, \p(d|\psi)\,\delta(s-s(\psi))\, \p(\psi) = \p(d|s)\, \p(s)$ and $\p(d)=\int \D s\, \p(d,s)$. The second equality in \eq{eq:psd} is just a renaming of the numerator and denominator of the first fraction, which highlights the connection to statistical mechanics. Thus we define the information Hamiltonian
\begin{equation}
 H(d,s) = - \log \p(d,s),
\end{equation}
the partition function including a moment generating source term $J$
\begin{equation}\label{eq:Zbeta}
 Z_\beta(d,J) = \int \D s\, e^{-\beta \, (H(d,s) + J^\dagger s)},
\end{equation}
and the inverse temperature $\beta = 1/T$ as usual in statistical mechanics. Here $s^\dagger$ is the transposed and complex conjugated signal vector $s$, leading to a scalar product 
$j^\dagger s = \int_V dx \,\bar{j}_x s_x$.
The ad-hoc notion of temperature is as in standard simulated annealing practice. It permits to narrow (for $T<1$) or widen (for $T>1$) the explored phase space region with respect to the one of the joint PDF and therefore is a useful auxiliary parameter.
We show  in Sect.~\ref{sec:tempered-posterior} that the well known thermodynamical equipartition theorem holds:
\begin{equation} \label{eq:equi-partition}
	\expectWRT{H(s,\,d)}{(s|d)} - H(m,\,d) \approx  \frac{1}{2}\, N_\mathrm{dgf}\,T.
\end{equation}
where $N_\mathrm{dgf}$ is the number of degrees of freedom and $m$ is the mean signal field as defined below in \eq{eq:mean-field}, which defines the ground state energy.
This relation can e.g.~be used to check the correctness of an implementation of a signal phase-space sampling algorithm.

\subsection{Maximum a posteriori}
\label{sec:MAP}

The first guess for a suitable energy to be minimized to obtain the information state might be the Hamiltonian. Minimizing the Hamiltonian with respect to $s$, while keeping $d$ at their observed values, is equivalent to maximizing the joint probability $\p(d,s)$ and also the posterior $\p(s|d)$. The classical field emerging from this is called the MAP signal reconstruction in signal processing. For a detailed discussion of the usage of the MAP principle in IFT see \cite{Lemm2003}. The MAP field is often a very good approximation of the mean field
\begin{equation}\label{eq:mean-field}
 m = \expectWRT{s}{(s|d)} \equiv \int \D s \, s\, \p(s|d),
\end{equation}
which is the optimal estimator of the signal in a statistical $L^2$ error norm sense:
\begin{equation}
 m = \mathrm{argmin}_{\tilde{s}} \expectWRT{\int_V dx \,(s_x-\tilde{s}_x)^2}{(s|d)}.
\end{equation}
The MAP estimator on the other hand can be shown to optimize the statistical $L^0$ norm\footnote{The $L^0$ norm measures the amount of exact agreement via $\| f \|_{0} = \lim_{\varepsilon\rightarrow 0} \frac{1}{\varepsilon} \int\! dx \, \theta(f^2(x)-\varepsilon^2)$, with $\theta$ denoting the Heaviside function.}, the result of which may strongly deviate from the mean $m$, if the posterior is highly asymmetric around its maximum. Thus we can regard the MAP estimator as a good reference point, but not as the solution we are seeking in general. It is, however, accurate (in the $L^2$ error norm sense) in case the posterior around its maximum is close to a Gaussian. In this case, the MAP field can easily be augmented with some uncertainty information from the Hessian of the Hamiltonian
\begin{equation}
 \mathcal{H} = \left. \frac{\delta^2 H(d,s)}{\delta s\, \delta s^\dagger}\right|_{s=m},
\end{equation}
as an approximation of the two point function of the signal uncertainty
\begin{equation}
 D \equiv  \expectWRT{(s-m)\,(s-m)^\dagger}{(s|d)}. 
\end{equation}
Thus we set $D\approx \mathcal{H}^{-1} $ in 
\begin{equation}\label{eq:psdg}
 \p(s|d)  \approx \tilde{\p}(s|d) = \G(s-m, D),
\end{equation}
where we introduced the Gaussian
\begin{equation}\label{eq:Gaussian}
 \G(\phi, D) \equiv \frac{1}{|2\pi\, D|^\frac{1}{2}} e^{-\frac{1}{2}\, \phi^\dagger D^{-1} \phi}.
\end{equation}
Unfortunately, the MAP estimator can perform suboptimally in cases where the Gaussian approximation does not hold, see e.g.  \cite{2010arXiv1002.2928E}.  

\subsection{Maximum Entropy}
\subsubsection{Image entropy}
\label{sec:MaxEnt}

Another quantity often extremized in image reconstruction problems is the so-called \textit{image entropy} (iE) 
\cite{
1978Natur.272..686G, 
1979MNRAS.187..145S, 
1980MNRAS.191...69B, 
1983CVGIP..23..113B, 
1983iimp.conf..267G, 
1984Natur.311..446S, 
1984Natur.312..381T, 
1984MNRAS.211..111S, 
1986JMOp...33..287B, 
gull1989, 
gullskilling,  
1998mebm.conf....1S}. 
In classical maximum (image) entropy (MiE here, usually ME) methods the iE is defined for a strictly positive signal via
\begin{equation}\label{eq:MaxEntEnt}
 S_\mathrm{iE}(s) = - \int_V dx\, s_x \,\log(s_x/\tilde{s}_x) \equiv - s^\dagger \log(s/\tilde{s}),
\end{equation}
where $\tilde{s}_x$ is the reference image, which is used to model some prior information. In the second equality we have defined the component-wise application of functions on fields, e.g. $(f(s))_x = f(s_x)$, which we use throughout this work.

We note, that the iE is actually not a physical entropy. Usually its usage is argued for by ad hoc assumptions on the distribution entropy of photon packages in the image plane, rather than being a well motivated description of the signal prior knowledge (or lack thereof). In the following we will reveal the implicitly assumed prior of MiE methods.

The data enter the MiE method in form of an image energy, which is ideally chosen to be the negative log-likelihood,
\begin{equation}
 E(d|s) = - \log\bigl(\p(d|s)\bigr),
\end{equation}
in order to ensure the best imprint of the data on the reconstruction.
The entropy is then maximized with the energy constraint given by minimizing
\begin{equation}\label{eq:EME}
 E_\mathrm{iE}(d,s) = E(d|s) - T\, S_\mathrm{iE}(s)
\end{equation}
with respect to $s$. Here $T$ is some adjustable temperature-like parameter, permitting us to choose the relative weight of image entropy and image energy. Low temperature means that the MiE map follows the data closely, high temperature that the map space wants to be more uniformly occupied by the signal reconstruction.

The prior information on the signal, $\p(s)$, does not enter the MiE formalism explicitly. Actually, an implicit prior can be identified, assuming that MiE is actually a MAP principle. In that case the implicitly assumed Hamiltonian is $H_\mathrm{iE}(d,s) \cong E_\mathrm{iE}(d,s)$, where $\cong$ denotes equality up to an irrelevant, since $s$-independent, additive  constant, and we find
\begin{equation}
\p_\mathrm{iE}(s) \propto e^{T\, S_\mathrm{iE}(s)} \propto  \prod_x  \,\left(\frac{s_x}{\tilde{s}_x}\right)^{-T\, s_x}.
\end{equation}
This is not a general prior, but a very specific PDF. Although there is some flexibility to adopt its functional form by choosing $\tilde{s}$, $T$, and the image space (pixel space, Fourier space, wavelet space, etc.) in which \eq{eq:MaxEntEnt} holds, $\p_\mathrm{iE}(s) $ can not be regarded as being generic. The MiE prior strongly suppresses large values in the MiE map. If a data feature can be either explained by a single map pixel exhibiting a peak value or by several pixels dividing that value among themselves, MiE will usually prefer the second option, leading to blurred reconstructed images.

We conclude, that the term \textit{maximum entropy} commonly used in image reconstruction is very misleading. A more accurate term would be \textit{minimal dynamical range}, since the implicitly assumed prior states that pixels carrying larger than average signal $s_x$ are extremely unlikely.

\subsubsection{Physical entropy}

A physical entropy should measure the distribution spread of a PDF using a phase space integral over its phase space. 
In fact, the latter is given by the Boltzmann entropy as given by the negative Shannon information,
\begin{equation}\label{eq:SB}
 S_\mathrm{B} = - \int \D s \, \p(s|d) \, \log \p(s|d),
\end{equation}
which is a functional of the signal posterior, $S_\mathrm{B}= S_\mathrm{B}[\p(s|d)] $, and not of the signal map. Inserting \eq{eq:psd} yields
\begin{equation}\label{eq:E-F}
 S_\mathrm{B} = \expectWRT{H(d,s)}{(s|d)} +  \log\, Z_1(d,0) = U - F,
\end{equation}
where we introduced the internal energy $U=  \expectWRT{H(d,s)}{(s|d)}$ and the Helmholtz free energy $F= F_1(d,0)$ with
\begin{equation}\label{eq:defF}
 F_\beta(d,J) = - \frac{1}{\beta} \,\log\, Z_\beta(d,J).
\end{equation}
The fully $J$-dependent Helmholtz free energy provides the field expectation value via
\begin{equation}\label{eq:m=dFdJ}
 m= \expectWRT{s}{(s|d)}= \left. \frac{\partial F_\beta(d,J)}{\partial J}\right|_{\beta=1, J=0}.
\end{equation}
The entropy is also given in terms of the free energy via
\begin{equation}
 S_\mathrm{B} = \left. \frac{\partial F_\beta(d,J)}{\partial \beta}\right|_{\beta=1, J=0}.
\end{equation}
The entropy as well as the free energy are functionals of the posterior and not of the signal. Maximizing or minimizing them does not provide a signal estimator, but singles out a PDF. If we restrict the space of PDFs to the ones we can handle analytically, namely Gaussians as given in \eq{eq:psdg} and \eq{eq:Gaussian}, we might obtain a suitable approximation scheme to the full field theoretical inference problem.

Maximizing the entropy alone does not lead to a suitable algorithm, since the maximal entropy state is that of complete lack of knowledge, with a uniform probability for every signal possibility. The internal energy, however, favors knowledge states close to the posterior maximum and would return the MAP solution if extremized alone. Thus the right combination of entropy and internal energy is to be extremized. 
We would expect a free energy of the form $U- T\, S_\mathrm{B}$ to be this function, in analogy to the energy \eq{eq:EME} used in MiE methods.
Thermodynamics teaches us that the Gibbs free energy is the quantity to be minimized (which is identical to the Helmholtz free energy in case $J=0$). Since we are going to calculate this for an approximation of the real PDF, it is necessary to go through the derivation in order to make sure we do this in the right fashion and understand all implications.

\section{Thermodynamical inference}\label{sec:thermoinf}
\subsection{Tempered Posterior}\label{sec:tempered-posterior}

In order to take full advantage of the existing thermodynamical machinery we want to construct the Gibbs free energy for information problems. 
To this end, we introduce a temperature and a source function into the PDF of the signal posterior as suggested by the definition of the partition function \eq{eq:Zbeta} by defining
\begin{equation}\label{eq:P(s|d,T,J)}
 \p(s|d,T,J) = \frac{e^{-\beta\,(H(s,d)+J^\dagger s)}}{Z_\beta(d,J)} = \frac{(\p(d,s)\,e^{-J^\dagger s})^\beta}{\int \mathcal{D}s'\,(\p(d,s')\,e^{-J^\dagger s'})^\beta}.
\end{equation}
With the temperature we can broaden (for $T>1$) or narrow (for $T<1$) the posterior. Three temperature values are of special importance, namely $T=0$, which modifies the PDF into a delta peak located at the posterior maximum, $T=1$, which returns the original posterior, and $T=\infty$, leading to the maximum entropy state of an uniform PDF. The source function $J$ permits us to shift the mean of the PDF to any possible signal configuration $m=m(d,T,J)$. 

The modified PDF will be approximated by a Gaussian with identical mean and variance: 
\begin{equation}\label{eq:myG}
 \p(s|d,T,J) \approx \G(s-m, D) = \tilde{\p}(s|m,D),
\end{equation}
where also $D=D(d,T,J)$. 

We will see, that the width $D$ of this Gaussian approximation of the PDF increases with increasing temperature.
At low temperature ($T \ll 1$) the center of the PDF is probed and modeled, while at large temperatures ($T\gg 1$)
the focus is on its asymptotic tails. 
Since the Gaussian in \eq{eq:myG} is an approximation, it is not even guaranteed that $T=1$ provides the best recipe for signal reconstruction. E.g. in \cite{2009PhRvD..80j5005E} a case is shown, where signal reconstruction using $T=0.5$ slightly outperforms both, $T=0$ and $T=1$. Since working at multiple temperatures can reveal different aspects of the same non-Gaussian PDF (i.e.~its central or asymptotic behavior), the question appears how the differently retrieved Gaussian approximations can be combined into a single and more accurate representation of the original PDF.
This will be addressed in Sect.~\ref{sec:infosynth}.
For the moment we approximate our posterior by a single Gaussian as in \eq{eq:myG}.

In this case, the partition function can be calculated explicitly and reads
\begin{displaymath}
	\tilde{Z}_\beta (d,\,J) = \left|\frac{2\pi}{\beta}  D \right|^{1/2}\!\!\!\!\!\! \exp\left({\frac{ J^\dagger D \,J}{2\beta} + J^\dagger m - \beta H(m,\,d)}\right)\!\!.
\end{displaymath}
With standard thermodynamics procedure we calculate
\begin{equation}
	\expectWRT{H}{(s|d)}\approx -\left.\frac{\delta}{\delta \beta}\tilde{Z}_\beta(d,\,J)\right|_{J=0} = \frac{N_\mathrm{dgf}}{2}\, T + H(m,\,d)
\end{equation}
where $N_\mathrm{dgf}$ is the dimension of the signal vector. This result is the re-phrased equipartition theorem \eq{eq:equi-partition} from classical thermodynamics and further motivates the notion of temperature in IFT.

\subsection{Internal, Helmholtz and Gibbs energy}
The next step is to calculate the Helmholtz free energy. In case it can be calculated explicitly from \eq{eq:defF}, the inference problem is basically solved, since any (connected) moment of the signal posterior can directly be calculated  from it by taking derivatives with respect to the moment generating function $J$, e.g. see \eq{eq:m=dFdJ}. This will, however, only be the case for a very restricted class of Hamiltonians, like the free ones, which are only quadratic in $s$. In the more interesting case the Helmholtz free energy can not be calculated explicitly, we can use the thermodynamical relation of the Helmholtz free energy with the internal energy and entropy. 

First, we note that the internal energy of the modified posterior is given by
\begin{eqnarray}
\label{eq: U(d,T,J)}
 U(d,T,J) &=& \expectWRT{H(s,d)}{(s|d,T,J)} \nonumber\\
&\approx &\expectWRT{H(s,d)}{(s|m,D)} = \tilde{U}(d,m,D),
\end{eqnarray}
where $m$ and $D$ are still functions of $d$, $T$, and $J$. The average in the second line has to be understood to be performed over a Gaussian with mean $m$ and dispersion $D$: $\langle f(s) \rangle_{(s|m,D)} = \int\mathcal{D}s\,f(s)\,\G(s-m,D)$.

Further, we need to calculate the entropy for the modified PDF, which for a Gaussian depends only on $D$:
\begin{equation}
S_\mathrm{B}[\G(s-m,D)] = \frac{1}{2} \, \Tr\bigl(1+\log(2\pi\,D)\bigr) = \tilde{S}_\mathrm{B}(D).
\end{equation}
For the full modified posterior, \eq{eq:P(s|d,T,J)}, the entropy is calculated via \eq{eq:SB} to be
\begin{equation}\label{eq:SBmodPDF}
 S_\mathrm{B} = \beta \,\left(U + J^\dagger m - F \right),
\end{equation}
where $m = m(d,T,J) = \expectWRT{ s }{(s|d,T,J)}$, $U$ is given by \eq{eq: U(d,T,J)}, and 
$F$ by \eq{eq:defF}. Solving \eq{eq:SBmodPDF} for the Helmholtz free energy yields
\begin{equation}
 F_\beta(d,J) = U - T \, S_\mathrm{B} + J^\dagger m.
\end{equation}
This expresses the Helmholtz free energy in terms of internal energy and entropy. Unfortunately, this expression contains the term $J^\dagger m$, where $m$ depends on $J$ implicitly through \eq{eq:m=dFdJ}. 
In order to get rid of this term, we
Legendre transform with respect to $J$ and thereby use \eq{eq:m=dFdJ}, which provides us with the Gibbs free energy 
\begin{equation}
 G_\beta(d,m) = F -  J^\dagger \frac{\delta F}{\delta J}= U - T \, S_\mathrm{B}.
\end{equation}
The Gibbs energy depends solely on $m$ and not on $J$. It can be constructed approximatively, in case approximations of the internal energy and the entropy are available. For our Gaussian approximation of the modified posterior we therefore write
\begin{equation}
\label{eq:Gibbs}
 \tilde{G}_\beta(d,m,D) = \tilde{U}(d,m,D) - T \,\tilde{S}_\mathrm{B}(D).
\end{equation}
We know from thermodynamics that the minimum of the Gibbs free energy with respect to variations in $m$ provides the expectation value $\expectWRT{s}{(s|d)}$ of our field:
\begin{equation}\label{eq:dGdm2} 
\left. \frac{\delta G(d,m,D)}{\delta m}\right|_{m=\expectWRT{s}{(s|d)}} \!\!\!\!\!\!\!\!\!\!\!\! \!\!\!\!\!\!\!\!\!\!\!\!= 0
\end{equation}
Thus, the Gibbs energy is the information energy we were looking for in the introduction.

\label{sec:generic formalism}
Minimizing the Gibbs free energy for a Gaussian PDF with respect to $m$ yields 
\begin{eqnarray}
0&=& \frac{\delta \tilde{G}}{\delta m} = \int \D s\,  H(d,s)\, \frac{\delta \, \G(s-m,D)}{\delta m}
\nonumber\\
&=& - D^{-1} \, \expectWRT{ \phi\, H_m(d,\phi)  }{(\phi|D)},
\end{eqnarray}
with $H_m(d,\phi)=H(d,m+\phi)$,
which implies
\begin{equation}
 m = \frac{\expectWRT{ s\, H(d,s)  }{(s|m,D)}}{\expectWRT{ H(d,s)  }{(s|m,D)}} = 
\frac{\expectWRT{ s\, H(d,s)  }{(s|m,D)}}{\tilde{U}(m,D)}.
\end{equation}
The optimal map is therefore the first signal moment of the full Hamiltonian weighted with the approximating Gaussian.

Thermodynamics teaches us further that the propagator, the uncertainty dispersion of the field, is provided by the second derivative of the Gibbs free energy around this location, thanks to the well known relation
\begin{equation}\label{eq:GFD}
 \left. \left(  \frac{\delta^2 G}{\delta m\,\delta m^\dagger} \right)^{-1}\right|_{m=\expectWRT{s}{(s|d)}} \!\!\!\!\!\!\!\!\!\!\!\!
= - \left. \frac{\delta^2 F}{\delta J\,\delta J^\dagger}  \right|_{J=0} =\beta\, D.
\end{equation}
This relation closes the set of equations by providing $D$. Evaluating \eq{eq:GFD} with our approximate Gibbs energy \eq{eq:Gibbs} and using \eq{eq:dGdm2} yields
\begin{eqnarray}
 T\,D^{-1} &=&  \left. \frac{\delta^2 \tilde{G}}{\delta m\,\delta m^\dagger}\right|_{m=\expectWRT{s}{(s|d)}}
=  -D^{-1}\,\tilde{U}(d,m, D)
\nonumber\\
&+& D^{-1}\,\expectWRT{ \phi\,\phi^\dagger H_m(d,\phi) }{(\phi|D)}\, D^{-1}.\nonumber
\end{eqnarray}
Thus the propagator is the second moment of the Gaussian weighted Hamiltonian,
\begin{equation}\label{eq:Dfromphiphi}
 D = \frac{\expectWRT{ \phi\,\phi^\dagger H_m(\phi)  }{(\phi| D)}}{\tilde{U}(d,m,D) +T}.
\end{equation}
This equation seems to suggest that the propagator evaluated at higher temperature is narrower, since $T$ appears in the denominator. However, the opposite is the case due to the presence of $D$ in all terms, as a test with a free Hamiltonian will show in \eq{eq:D=TD*}.

\subsection{Cross information}

The Gibbs free energy at $T=1$ is directly related to the cross information between the posterior and its Gaussian approximation. The cross information (or negative relative entropy) of a PDF $\tilde{\p}$ with respect to another one $\p$ is measured by the so called Kullback-Leibler divergence  \cite{Kullback1951}:
\begin{equation}\label{eq:FreeeKL}
\mathrm{d}_\mathrm{KL}[\tilde{\p},\p]
=
\int\D s \, \tilde{\p}(s|d)\, \log\left( \frac{\tilde{\p }(s|d)}{\p (s|d)}\right).
\end{equation}
The Kullback-Leibler divergence characterizes the distance between a surrogate and target PDF in an information theoretical sense. It is an asymmetric distance measure, reflecting that the roles of the two involved PDF differ. The equivalence of Gibbs free energy and cross information with respect to inference problems can easily be seen:
\begin{eqnarray}
 \tilde{G}(m,D) &=&  \expectWRT{ H(d,s) + \log(\G(s-m,D))}{(s|m,D)} \nonumber\\
&=&  \int\D s\, \G(s-m,D)\, \log\left( \frac{\G(s-m,D)}{\p(s,d)}\right) \nonumber\\
& \cong&  \int\D s\, \G(s-m,D)\, \log\left( \frac{\G(s-m,D)}{\p(s|d)}\right)\!\! \;\;\,\,\,\,\,\,\nonumber\\
&=& \mathrm{d}_\mathrm{KL}[\tilde{\p},\p].
\end{eqnarray}
In the second last step we added the term $\log\p(d)$, which is irrelevant here, since $m$- and $D$-independent, and in the last step we introduced the  Kullback-Leibler divergence between posterior $\p(s|d)$ and its Gaussian surrogate $\tilde{\p}(s|d) = \G(s-m,D)$. Minimal Gibbs free energy therefore seems to corresponds to minimal Kullback-Leibler divergence, and therefore to maximal cross information of the surrogate with the exact posterior.

However, we have only minimized the Gibbs free energy so far with respect to $m$, the mean field, degrees of freedom of our Gaussian, not with respect to the ones parameterizing the uncertainty dispersion $D$. We have determined this using the thermodynamical relation \eq{eq:dGdm2}. If we want that our surrogate PDF has maximal cross information with the posterior with respect to all degrees of freedom of our Gaussian, we also have to minimizing the Gibbs energy with respect to $D$.
A short calculation shows that this actually yields a result which is equivalent to the thermodynamical relation \eq{eq:GFD}:
\begin{eqnarray}
 0  &=&\frac{\delta \tilde{G}}{\delta D} = \int \D\phi\,  H_m(d,\phi)\, \frac{\delta \, \G(\phi,D)}{\delta D} - T \, 
 \frac{\delta \, \tilde{S}_\mathrm{B}(D)}{\delta D}\nonumber\\
&=& \frac{D^{-1}}{2}  \left[\expectWRT{ \phi\,\phi^\dagger H_m(d,\phi) }{(\phi|D)} 
- D\,\bigl(\tilde{U}(m, D) + T\bigr)
\right] D^{-1}, \nonumber
\end{eqnarray}
from which also \eq{eq:Dfromphiphi} follows. 
Thus, we can regard both, the map $m$ and its uncertainty covariance $D$, as parameters for which the Gibbs energy should be minimized. We will refer to this as the maximal cross information principle. 

We further note that the maximal cross information principle also holds if the Gaussian is replaced by some other model function, $G[\tilde{\p}(s|d)] \cong  \mathrm{d}_\mathrm{KL}[\tilde{\p},\p]$,
a property we will use later in Sect.\ \ref{sec:infosynth}. 

Note, that the minimal cross information and the thermodynamical relations yield exactly the same results for $m$ and $D$ only if $\tilde{G}(m,D)$ is calculated exactly. In case there are approximations involved, the resulting algorithms differ slightly, and this difference can be used to monitor the impact of the approximation made. In the following, we use the minimal cross information principle for our examples.

\subsection{Calculating the internal energy}

In order to calculate the approximative Gibbs energy, we need to estimate the internal energy, for which we have to specify the exact Hamiltonian. We assume that it can be Taylor-Fr\'echet expanded as
\begin{equation}
\label{eq:TaylorH}
 H(d,s)= \sum_{n=0}^\infty \, \frac{1}{n!}\,
\underbrace{\Lambda^{(n)}_{x_1 \ldots x_n}\, s_{x_1}\! \cdots s_{x_n}}_{ \Lambda^{(n)}(s, \ldots s)},
\end{equation}
where repeated coordinates are thought to be integrated or summed over. The approximative internal energy is then
\begin{eqnarray}
 \tilde{U}(m,D) &=& U[\tilde{\p}(s|d)] = \int\D s\, H(d,s)  \, \tilde{\p}(s|d) \nonumber\\
%
%
&=& \sum_{n=0}^\infty \, \frac{1}{n!}\, \expectWRT{ \Lambda^{(n)}(s, \ldots s)}{(s|m,D)}
.
\end{eqnarray}
The Gaussian $n$-point correlation functions in this equation can actually be calculated analytically.
For this, we again use the shifted field $\phi = s-m$, which has the Hamiltonian
\begin{eqnarray}
  H_m(d,\phi) &=& \sum_{n=0}^\infty \, \frac{1}{n!}\, \Lambda_m^{(n)}(\phi, \ldots \phi),
\,\,\mathrm{with} \\
\Lambda_m^{(n)} \, (\phi, \ldots \phi) &=& \sum_{k=0}^\infty \, \frac{1}{k!}\, \Lambda^{(n+k)}(\underbrace{\phi, \ldots \phi}_n, \underbrace{m, \ldots m}_k) \nonumber.
\end{eqnarray}
We assume that the interaction coefficients $\Lambda^{(n)}_{x_1\ldots x_n}$  are symmetric with respect to index permutations, since they resulted from a Taylor-Fr\'echet expansion.

The internal energy can then be calculated via the Wick theorem and the fact that all odd moments of $\phi$ vanish:
\begin{eqnarray}\label{eq:U(m,D)}
 \tilde{U}(m, D) &=& \sum_{n=0}^\infty \, \frac{1}{n!}\, \expectWRT{ \Lambda_m^{(n)}(\phi, \ldots \phi)}{(\phi|D)}\\
&=& \sum_{n=0}^\infty \, \frac{1}{2^n\, n!}\, \Lambda_m^{(2n)}(\overbrace{D\otimes \cdots D}^n) \nonumber\\
&=&
\sum_{n,k=0}^\infty \, \frac{\Lambda^{(2n+k)}(\overbrace{D\otimes \cdots D}^{n} \otimes \overbrace{m\otimes \cdots m}^k) }{2^n\,n!\,k!}\nonumber.
\end{eqnarray}
Here, we defined the symmetrized tensor
product $\bigl(T\otimes T'\bigr)_{x_1\ldots x_n}\equiv \sum_{\pi\in
S_n}\frac{1}{n!}T_{x_{\pi(1)}\ldots x_{\pi(k)}}\cdot T'_{x_{\pi(k+1)}\ldots
x_{\pi(n)}}$ by averaging over all permutations in $S_n$, the symmetric group.

Having obtained the internal energy with \eq{eq:U(m,D)}, and entropy with \eq{eq:SBmodPDF} approximatively, we 
can construct the Gibbs free energy according to \eq{eq:Gibbs} which we use for our inference.
 
\subsection{Minimizing}

In order to get our optimal Gaussian approximation to the posterior, we have to minimize $ \tilde{G}_\beta (m,D)$ with respect to $m$ and $D$. Minimizing for $m$ is equivalent to minimizing the internal energy, since the entropy does not depend on $m$. This yields
\begin{eqnarray}\label{eq:mDeterminationAbstract} 
 0 &=& \frac{\delta \tilde{U}(m, D)}{\delta m}\\
&=&
\sum_{n,k=0}^\infty \!\!\!\! \frac{\Lambda^{(2n+k+1)}(\overbrace{D\otimes \cdots D}^{n} \otimes \overbrace{m\otimes \cdots m}^{k}, \cdot) }{2^n\,n!\, k!},\!\!\!\!\!\!\!\!\!\!\!\!\nonumber.
\end{eqnarray}
which has to be solved for $m$ for any given $D$. The propagator derives from \eq{eq:GFD} or from
\begin{eqnarray}\label{eq:DDeterminationAbstract}
 0 &=& \frac{\delta \tilde{G}(m, D)}{\delta D} \; \Rightarrow \\
T\, D^{-1} &=& 
\sum_{n,k=0}^\infty \!\!\!\! \frac{\Lambda^{(2n+k+2)}(\cdot, \cdot, \overbrace{D\otimes \cdots D}^{n} \otimes \overbrace{m\otimes \cdots m}^{k}) }{2^n\,n!\, k!} \!\!\!\!\!\!\!\!\!\!\!\!\nonumber.
\end{eqnarray}
which also depends on $m$. Thus, \eq{eq:mDeterminationAbstract} and \eq{eq:DDeterminationAbstract} have to be solved simultaneously. 

A simple example should be in order. The simplest case is that of the original Hamiltonian being quadratic. The approximated one should then match this exactly. A quadratic or free Hamiltonian is equivalent to a Gaussian posterior, $\p(s|d)= \G(s-m_*,D_*)$. We get
\begin{eqnarray}
\label{eq:freeH}
 H(d,s) &\cong& \frac{1}{2}\, (s-m_*)^\dagger D_*^{-1} (s-m_*)  \nonumber\\
&\cong&  \Lambda^{(1)}_{x} \,s_{x} +  \frac{1}{2}\, \Lambda^{(2)}_{xy} \,s_{x} s_{y}\,\;\; \mbox{with}\\
\Lambda^{(1)} &=& - D_*^{-1}m_*, \;\;\mbox{and}\nonumber\\
\Lambda^{(2)} &=&  D_*^{-1}.\nonumber
\end{eqnarray}
Inserting this into \eq{eq:mDeterminationAbstract} and \eq{eq:DDeterminationAbstract} yields
\begin{eqnarray}
 0 &=& \Lambda^{(1)}(\cdot) +  \Lambda^{(2)} (m,\cdot)  = D_*^{-1} (m-m_*)\nonumber\\
    &\Rightarrow& m=m_*,\\
T\, D^{-1} &=&  \Lambda^{(2)}(\cdot, \cdot)   =  D_*^{-1}  \nonumber\\
    &\Rightarrow& D=T\, D_*, \label{eq:D=TD*}
\end{eqnarray}
which indeed recovers the original coefficients for $T=1$, and a narrower or wider uncertainty dispersion for $T<1$ or $T>1$, respectively. In the following, we will see that also in case of interacting Hamiltonians the minimal free energy principle provides the correct results. We show this by reproducing (and extending) signal estimators derived previously in IFT using renormalization techniques.

\section{Application examples}\label{sec:application}
\subsection{Poissonian log-normal data}
\label{sec:plndata} 
\subsubsection{Separable case}

Many inference problems have to deal with Poissonian noise, like X-ray and $\gamma$-ray astronomy as well as reconstruction of the cosmic large-scale structure from galaxy counts. Let us assume that the mean count rate $\lambda$ of photons or galaxies is proportional to an exponentiated Gaussian random field $s$ with covariance $S=\expectWRT{ s\, s^\dagger }{(s)}$ according to
\begin{equation}
 \lambda(s) = \kappa\,e^{b\,s}\,.
\end{equation}
Here, $\kappa$ is the expected counts for $s=0$, which may depend on the spatial position. The scalar $b$ permits us to change conveniently the strength of the non-linearity of the problem without changing the signal statistics. This log-normal model for the cosmic large-scale structures as an approximative description is actually supported  observationally \cite{1934ApJ....79....8H, 2009MNRAS.400..183K} and theoretically \cite{Layzer1956,1991MNRAS.248....1C, 1995MNRAS.277..933S, Kayo2001, 2001PASP..113.1009V,2009ApJ...698L..90N}.

As a starting point, we assume a local response, so that the Poisson statistics for the actual counts $d_x$ at location $x$ are
\begin{equation}
 P(d_x| \lambda_x) = \frac{\lambda_x^{d_x}}{d_x!}\, e^{-\lambda_x},
\end{equation}
and the full likelihood is well separable into local ones:
\begin{equation}\label{eq:separablepoissonlike} 
 P(d|s) = \prod_x  P(d_x| \lambda_x(s_x)).
\end{equation}
The corresponding Hamiltonian was shown in \cite{2009PhRvD..80j5005E} to be
\begin{eqnarray}
\label{eq:sepPoissonHamilton}
 H(d,s) &\cong& \frac{1}{2} s^\dagger S^{-1} s - d^\dagger b\, s + \kappa^\dagger e^{b\,s}.
\end{eqnarray}
Reconstruction methods for this data model were developed by \cite{2009PhRvD..80j5005E, 
2010MNRAS.403..589K, 2009arXiv0911.2496J, 2009arXiv0911.2498J}.

The internal energy of our Gaussian approximation can be calculated analytically,
\begin{eqnarray}
 \tilde{U}(m,D) &\cong& \frac{1}{2} m^\dagger S^{-1} m + \frac{1}{2}\Tr(D\, S^{-1}) - d^\dagger b\,m \nonumber\\
&+& \kappa^\dagger e^{b\,m + \frac{b^2}{2}\widehat{D}} 
\end{eqnarray}
where $\widehat{D}$ denotes the vector of diagonal elements of $D$.

Minimizing $\tilde{G}(m,D)= \tilde{U}(m,D)-T\,\tilde{S}_\mathrm{B}(D)$ with respect to $m$ and $D$ yields
\begin{eqnarray}\label{eq:LNPsmD}
m &=& S\, b\,\left(d - \kappa_{m + \frac{b}{2}\widehat{D}}\right), \;\;\mbox{and} \nonumber\\
D &=& T \,\left( S^{-1} + b^2\,\widehat{\kappa}_{m+ \frac{b}{2}\widehat{D}} \right)^{-1},
\end{eqnarray}
respectively. Here we have defined $\kappa_t = \kappa \exp(b\,t)$ and denote a diagonal matrix by putting a hat onto a vector of its diagonal elements $(\widehat{\lambda})_{xy} = \lambda_x\, \delta_{xy}$. This result is identical with the one found in \cite{2009PhRvD..80j5005E} using a lengthy renormalization calculation. 
There it was found by numerical experiment, that using $T=0.5$ in \eq{eq:LNPsmD} seems to produce slightly better results than $T=0$ and $T=1$.

\subsubsection{Entangled case}
\label{sec:logPoissonEntagled}

So far, we assumed that the response provides a one to one correspondence between locations in signal and data space. However, for most measurements this is not exactly true. X- and $\gamma$-ray telescopes typically exhibit point spread functions, which map a single signal space location onto several detectors, of which each detects events coming from several indistinguishable directions. 
Also galaxy redshifts do not provide accurate distance information, since redshift distortions and measurement errors lead to effective point spread functions. 

In the following, we generalize to the case of a known and fixed, but non-local measurement response. Fixed means, that the response is independent of the signal. This excludes the treatment of galaxy redshift distortions with this case (e.g. see \cite{2010arXiv1003.1311W} for this), but still includes photometric redshift errors of galaxy catalogs as well as X- and $\gamma$-ray telescope data. Such problems have been approached in the past via the MAP principle \cite{Hebert1992,Cornwell1985,Wang2008,Oh2009}.

The point spread function is modeled by the response matrix $R = (R_{ix})$ which describes how emissivity at location $x$ is expected to be observed in data channel $i$. The expected count rate is now
\begin{equation}
 \lambda(s) =R\,e^{b\,s},
\end{equation}
and the likelihood does not separate any more with respect to $x$
\begin{equation}
 P(d|s) = \prod_i  P(d_i| \lambda_i(s)),
\end{equation}
since $\lambda_i(s)$ entangles the signal from several locations, whereas in \eq{eq:separablepoissonlike} it depends only on the local signal value. We recover the former case for a diagonal response $R_{ix}= \kappa_x\,\delta_{ix}$.
The resulting Hamiltonian 
\begin{eqnarray}
 H(s|d) &\cong& \frac{1}{2} s^\dagger S^{-1} s + 1^\dagger R\, e^{b\,s} - d^\dagger \log( R\, e^{b\,s})
\end{eqnarray}
reduces to \eq{eq:sepPoissonHamilton} for $R$ being diagonal.

The internal energy of our surrogate Gaussian $\tilde{\p}(s|d) = \G(s-m,D)$ is then
\begin{eqnarray}
\tilde{U}(m,D) &=& \frac{1}{2} \, m^\dagger S^{-1} m +
 \frac{1}{2}\Tr(D\,S^{-1}) + 1^\dagger R\, e^{b\,m + \frac{b^2}{2} \,\widehat{D}} \nonumber\\
&-& \sum_i d_i \underbrace{\int \!\! \D\phi \, \log\left(R_i^\dagger \, e^{b\,(m +\phi)}\right) \, \G(\phi,D)}_{\mathrm{I}_i}.
\end{eqnarray}
This integral $I_i$ can not be calculated in closed from due to the logarithm in the integrand. We expand the logarithm around $R_i^\dagger e^m$, since we will see that this recovers the result of the separable case most easily for $R$ being diagonal. We get
\begin{eqnarray}
 \mathrm{I}_i &=& \log\left( R_i^\dagger e^{b\,m} \right) +  
\expectWRT{ \log\left( \frac{R_i^\dagger \, e^{b\,(m +\phi)}}{R_i^\dagger e^{b\,m}} \right) }{(\phi|D)}.
\end{eqnarray}
In case $R$ is diagonal, the first term reduces to $b\,m+ \log R_i$, the second vanishes as $\expectWRT{ \log (\exp(b\, \phi)) }{(\phi|D)} = \expectWRT{ b\, \phi }{(\phi|D)} = 0$, and we recover the Hamiltonian of the separable case. 

In the general case of an entangling response we Taylor expand the logarithm of the second term
\begin{eqnarray}
 \mathrm{I}_i &=& \log\left( R_i^\dagger e^{b\,m} \right) \nonumber\\
&-&   \sum_{n=1}^{\infty} \frac{(-1)^n}{n}
\underbrace{
\expectWRT{ \left(r_i^\dagger   e^{b\,\phi} - 1 \right)^n }{(\phi|D)}
}_{\mathrm{II}_{i\,n}},\;\;\mbox{with} \\
r_{i} &=& \frac{R_{i}e^{b\,m}}{R_i^\dagger e^{b\,m}}\;\mbox{or}\; r_i(x)= \frac{R_{i}(x)\, e^{b\,m(x)} }{\int \! dx'\, R_{i}(x')\,e^{b\,m(x')}}.\nonumber
\end{eqnarray}
We note that $r_i^\dagger 1 = \int dx\, r_{ix} =1$ by construction. 

The expansion coefficients $\mathrm{II}_{i\,n}$ can be worked out one by one. We provide here the first few, namely
\begin{eqnarray}
 \mathrm{II}_{i\,1} &=& 
r_i^\dagger    e^{\frac{1}{2}b^2 \widehat{D}} - 1,\nonumber\\
 \mathrm{II}_{i\,2} 
& = & r_{ix} r_{iy}  \, e^{\frac{1}{2}b^2 (D_{xx} + D_{yy} + 2D_{xy})} -2 r_i^\dagger   \, e^{\frac{1}{2}b^2 \widehat{D}} + 1,\nonumber\\
 \mathrm{II}_{i\,3} 
& = & r_{ix} r_{iy}  r_{iz}\, \exp\left( \frac{b^2}{2} \!\!\!\!\!\! \sum_{a,b\in\{ x,y,z\} } \!\!\!\!\!\! D_{ab}\right) 
\nonumber\\
&-& 3  \, r_{ix} r_{iy}  \, \exp\left( \frac{b^2}{2} \!\!\!\! \sum_{a,b\in\{ x,y\} } \!\!\!\! D_{ab}\right) \nonumber\\
&+& 3 \, r_i^\dagger e^{\frac{1}{2}b^2 \widehat{D}} - 1.
\end{eqnarray}
These coefficients stay small if $b^2\,D \ll 1$, which means that the expansion can be truncated if the signal is known within a few ten percent or if non-Gaussianity is small. Large uncertainties in the signal strength do not necessarily lead to large coefficients if they are located at positions without instrumental sensitivity ($R_{ix}$ small) or much lower expected count rates ($m_x$ small). In both cases mostly prior information and extrapolation from regions with more informative data will determine the solution at such locations.

In case some of these coefficients are large, substantial signal uncertainty at the locations to which they are sensitive must be present. In this case an accurate reconstruction for these locations can not be expected. Thus, if we simplify the Hamiltonian by dropping such terms, even if they are relatively large, the quality of the reconstruction will not suffer too much since only regions are affected, which are poorly constrained by the data anyway. Therefore, truncating the expansion should already provide usable algorithms.

\subsubsection{Zeroth order solution}
\label{sec:0ordsol}

To zeroth order, we ignore all $\mathrm{II}_{i\,n}$-terms and find for the approximative free energy
\begin{eqnarray}
\tilde{G}(m,D) &\approx& \frac{1}{2} \, m^\dagger S^{-1} m +
 \frac{1}{2}\Tr(D\,S^{-1})\nonumber\\
&+& \sum_i \left[ R_i^\dagger e^{b\,m + \frac{b^2}{2} \,\widehat{D}} - d_i\, \log\left( R_i^\dagger e^{b\,m}\right)\right] \nonumber\\
&-& \frac{T}{2} \Tr\left(1+\log(2\pi\,D)\right). 
\end{eqnarray}
Minimizing this with respect to $m$ and $D$ yields
\begin{eqnarray}\label{eq:pln0thorder}
 m &=& S\,  b\, \sum_i R_i\, e^{b\,m}\, \left( \frac{d_i}{R_i^\dagger e^{b\,m}} - e^{\frac{1}{2}b^2\,\widehat{D}}\right) \nonumber\\
&=& S\, b\, \left(d^\dagger r - \kappa'({m+b\,\widehat{D}/2})\right), \; \mbox{and} \nonumber\\
D &=& T \,\left( S^{-1} + b^2\,\widehat{\kappa}'({m+ b\,\widehat{D}/2}) \right)^{-1}, \;\;\mbox{with}\nonumber\\
\kappa'(t) &=&
\sum_i \,R_{i}\,e^{b\,t}.
\end{eqnarray}
This is very similar to \eq{eq:LNPsmD} and reduces to it for a diagonal response. 

\subsubsection{First order correction}
\label{sec:1stordercor}

First order corrections are included by keeping the $\mathrm{II}_{i1}$-term in the approximative free energy, but ignoring higher terms. The resulting equations are
\begin{eqnarray}\label{eq:pln1storder}
 m &=& S\, b\, \left(\sum_i d_i \left( 1 + r_i^\dagger e^{\frac{b^2}{2} \widehat{D}} \right) r_i  - \kappa''({m+b\,\widehat{D}/2})\right)\nonumber\\
D &=& T \,\left( S^{-1} + b^2\,\widehat{\kappa}''({m+ b\,\widehat{D}/2}) \right)^{-1}, \;\;\mbox{with}\nonumber\\
\kappa''(t) &=&
\sum_i \,R_{i}\,e^{b\,t} \left( 1+ \frac{d_i}{R_i^\dagger e^{b\,m}}\right).
\end{eqnarray}
This is a slight modification with respect to \eq{eq:pln0thorder} in two aspects. The map changes a bit, but the sign of the changes depends on the details of the point spread function, since there are two new terms of similar order, but with opposite signs. The uncertainty variance is reduced, since the term added to the inverse propagator is always positive. 

\subsubsection{Observation with background}
\label{sec:obswithbackground}

The observation may suffer from a background, events in data space, which do not contribute to our signal knowledge. For example $\gamma$-ray astronomy has to suppress cosmic ray events as much as possible, since charged particles do not point back to the same sources as neutral photons due to cosmic magnetic fields. Fortunately, cosmic rays have different signatures in data space due to the differences in hadronic and electromagnetic interactions. However, not for all measured events is the distinction clearly cut and we have to use prior knowledge to suppress the background events. 

Therefore we should extend our formalism to also take such unwanted backgrounds into account. Actually a reinterpretation of the above formula will do. We extend our signal space by the quantity $f$ determining the logarithm of the background count rate, $s \rightarrow s'= (s,f)$. $f_z$ might be a field over the same physical space as $s_x$, or just a single number as a total isotropic cosmic ray flux. In any case, the $x-$ and $z-$coordinates are regarded to be over different spaces, or distinct areas of the joint space over which $f$ and $s$ live.
The joint covariance reads
\begin{equation}
 S' =
\left(
\begin{array}{ll}
S & 0 \\ 
0 & F
\end{array}
\right)
\end{equation}
due to the independence of signal and background. Here, $F=\expectWRT{ f\,f^\dagger }{(f)}$ is the log-background covariance.
The response $R \rightarrow R'$ has to be extended to map also the background space into the data space. 
Whether the response images of signal and background events in data space  are well separated or whether they overlap decides about the background discriminating power of the instrument.

The combined map and covariance of signal and log-background can now be obtained, e.g.~from \eq{eq:pln0thorder} or \eq{eq:pln1storder} with the appropriate replacements for $S,R,m,D\rightarrow S',R',m',D'$.
Our joint map can be split into a signal and log-background part $m' = (\tilde{s},\tilde{f})$.
Since we are usually not interested in the background properties, we marginalize over it. This is especially simple in the Gaussian approximation of our joint posterior $P(s'|d)\approx \G(s'-m',D')$, with $s'=(s,f)$,  $m' = (\tilde{s},\tilde{f})$,
\begin{eqnarray}
 m&\approx& \int \D s' \,s \, \G(s'-m',D') = \tilde{s},\;\mbox{and}\\
D_{xy}&\approx& \int \D s'\, (s-\tilde{s})_x\, (s-\tilde{s})_y\,\G(s'-m',D') = D'_{xy}.\nonumber
\end{eqnarray}
Although this does not look too different from the formula for the case without background, the effect of the background entered through the joint covariance matrix $D'$, which mixes the contribution from the signal and background events appropriately.

\subsection{Reconstruction without spectral knowledge}
\label{sec:rec. without spec. knowledge}
\subsubsection{Effective theory}

The reconstruction of the signal in the Poisson log-normal model in the previous section assumed that the signal covariance is known a priori. 
In case it is unknown, it has to be extracted from the same data used for the signal inference
\cite{1987AJ.....93..968R, 1992ApJ...398..169R, 1996ITSP...44.1469L, 1999ISPL....6..205S, 2000AJ....120.2163S}. However, the optimal way to do this was usually not derived from first principles, maybe except in \cite{1986WRR....22..499K, 2008ApJ...675.1304R, 1998ApJ...503..492S}. A rigorous approach to such problems is given by the computationally expensive Gibbs-sampling technique, which investigates the joint space of signal realizations and power spectra \cite{2004PhRvD..70h3511W, 2004ApJS..155..227E, 2004ApJ...609....1J, 2010MNRAS.406...60J}, which can then easily be marginalized over the power spectra to obtain a generic signal reconstruction. This problem was also addressed approximatively for the case of linear response data from a Gaussian signal subject to Gaussian noise using the MAP principle as well as by the help of parameter uncertainty renormalized estimation by \cite{2010arXiv1002.2928E}.  We re-address this problem here using the minimal free energy approach.

We assume the covariance $S = \expectWRT{ s\,s^\dagger}{(s)}$ of our Gaussian signal $s$ to be diagonal within some known function basis 
$O_{kx}$, e.g. the Fourier basis with $O_{kx}= e^{i\,k\,x}$. We model the power spectrum (in this basis) as being a linear combination of a number of positive basis functions $f_i(k)$ with disjoint supports (the spectral bands), so that
\begin{equation}
P_{s}(k) = \sum_i p_i f_i(k)
\end{equation}
is positive for all $k$ (all coefficients of $p=(p_i)_i$ are positive and the spectral bands cover the full $k$-space domain).
We define 
\begin{equation}
 (S_i)_{xy} = (O^\dagger \widehat{f}_i O)_{xy}= \overline{O_{k\,x}} \, f_i(k) \, O_{k\, y} 
\end{equation}
to be the $i$-th spectral band matrix  and $S_i^{-1}$ to be its
pseudo-inverse. Thus, we write our signal covariance as
\begin{equation}
 S = \sum_i p_i S_i,
\end{equation}
with $p=(p_i)$ the vector of unknown spectral parameters. We further assume that the individual signal-band amplitudes $p_i$ have an independent prior distribution,
\begin{equation}\label{eq:pall-prior}
 \p(p) = \prod_i \p(p_i), 
\end{equation}
with the individual priors being inverse-gamma distributions, power-laws with 
exponential low amplitude cutoff at $q_{i}$ :
\begin{equation}
\label{eq:p-prior}
\p(p_i) = \frac{1}{q_i\,\Gamma(\alpha_i -1)} \,\left(\frac{p_i}{q_i}\right)^{-\alpha_i} \, 
\exp \left(-\frac{q_i}{p_i}\right).
\end{equation}
For $\alpha_i \gg 1$ this is an informative prior, where  $q_i/\alpha_i$ determines the preferred value.
A non-informative prior would be given by Jeffreys prior with $\alpha_i = 1$ and $q_i=0$.\footnote{Since this would result in an improperly normalized prior, we understand this as $\alpha_i= 1+ \epsilon $, $q_i = \epsilon$, and $\lim_{\epsilon \rightarrow 0}$ at the end of the calculation.}

For a linear data model
\begin{equation}
 d= R\, s +n,
\end{equation}
with Gaussian noise with covariance $N = \expectWRT{ n\,n^\dagger}{(n)}$, the parameter marginalized effective Hamiltonian is according to \cite{2010arXiv1002.2928E} 
\begin{equation}\label{eq:MAPmapHamilton}
 H(d,s) \cong\frac{1}{2} \, s^\dagger M\, s - j^\dagger s + \sum_i  \gamma_i\, \log\left(q_i+ \frac{1}{2} \, s^\dagger S_i^{-1} s\right).
\end{equation}
Here $M=R^\dagger N^{-1} R$, $j=R^\dagger N^{-1} d$,  $\gamma_i = \alpha_i -1 + \varrho_i/2$, and $\varrho_i= \Tr[S_i^{-1}S_i]$ the number of spectral degrees of freedom within the band $i$.

\subsubsection{Free energy expansion}
\label{sec:purefeeee}

The internal energy of a Gaussian posterior-ansatz is then
\begin{eqnarray}
 \tilde{U}(m,D) &\cong& \frac{1}{2} \, m^\dagger M\, m + \frac{1}{2} \,\Tr(D\,M) - j^\dagger m \nonumber\\
&+& \sum_i  \gamma_i\,  \underbrace{ \expectWRT{ \log\left(q_i+ \frac{1}{2} \, s^\dagger S_i^{-1} s\right) }{(s|m,D)}}_{\mathrm{I}_i} .
\end{eqnarray}
Again we have to deal with a Gaussian average over a logarithm, which we expand as
\begin{eqnarray}
\mathrm{I}_i &=& \log(\tilde{q}_i) - \sum_{k=1}^{\infty} \frac{(-1)^k}{k\,(\tilde{q}_i )^k}
\underbrace{ \expectWRT{\left(q_i+ \frac{1}{2} \, s^\dagger S_i^{-1} s- \tilde{q}_i\right)^k }{\!\!\!(s|m,D)}}_{\mathrm{II}_{ik}}\!\!\!\!,\nonumber\\
&&\mbox{with}\;\;
 \tilde{q}_i = q_i+ \frac{1}{2} \, \Tr((m\, m^\dagger + \delta \, D) S_i^{-1} ).
\end{eqnarray}
Here we have introduced a parameter $\delta$  to be fixed soon. The first two expansion coefficients are
\begin{eqnarray}
 \mathrm{II}_{i1} &=& \frac{1}{2}\, (1-\delta) \Tr(D\, S_i^{-1})\nonumber\\
 \mathrm{II}_{i2} &=&   \mathrm{II}_{i1}^2
%
+ \Tr\left(\left(m\, m^\dagger +   \frac{1}{2}\,D\right) S_i^{-1}D\, S_i^{-1}\right).
\end{eqnarray}

\subsubsection{Zeroth order solution}
To zeroth order we find by minimizing the free energy while ignoring the $ \mathrm{II}$-corrections
\begin{eqnarray}
 m &=& D' \, j, \; D= T\,D',\;\mbox{and}\nonumber \\
D' &=& \left(M + \sum_i p_i^{-1} S_i^{-1} \right)^{-1}.
\end{eqnarray}
This means that the map is the Wiener filtered data, where the spectral coefficients are assumed to be
\begin{equation}\label{eq:pi0thOrder}
 p_i = \frac{\tilde{q}_i}{\gamma_i\, \delta} = \frac{1}{\gamma_i\, \delta} \left( q_i+ \frac{1}{2} \, \Tr((m\, m^\dagger + \delta \, D) S_i^{-1} )\right).
\end{equation}
For $\delta=0$ this yields $p_i=\infty$ and therefore $D=M^{-1}$ if $M$ is (pseudo)-invertible. The resulting filter provides a noise weighted deconvolution, however is unable to extrapolate into unobserved regions of the signal space. It is widely used for map making in the field of cosmic microwave background observations.
For $\delta=1$ we recover the critical estimator of \cite{2010arXiv1002.2928E}. Since there it was shown that the latter performs significantly better than the former, and also since $\mathrm{II}_{i1}=0$ and  $\mathrm{II}_{i2}$ is minimal for $\delta=1$, we adopt this in the following. For Jeffreys prior we find
\begin{eqnarray}\label{eq:p_i0}
  p_i &=& \frac{\Tr(B_i)}{\varrho_i},
\end{eqnarray}
with  $B_i=(m\,m^\dagger+D) S_i^{-1}$.

\subsubsection{Second order correction}
\label{sec:pure2ndorder}
Including higher order corrections should improve the reconstruction. The first order corrections vanish for $\delta=1$. The second order correction yields 
\begin{eqnarray}
m &=& D'\, j,\; 
D = T \left[ {D'}^{-1} - \sum_i \frac{\gamma_i}{\tilde{q}_i^2} \,  S_i^{-1} m\, m^\dagger S_i^{-1}\right]^{-1},
\nonumber\\
D' &=& \left[ M + \sum_i \frac{\gamma_i}{\tilde{q}_i} \, 
X_i \, S_i^{-1} \right]^{-1},\\
X_i &=& 1 + 
\frac{1}{\tilde{q}_i^2} \Tr\left((m\,m^\dagger + \frac{1}{2}\,D) \,S_i^{-1} D  \, S_i^{-1}\right) -  \frac{1}{\tilde{q}_i} \,S_i^{-1} D. \nonumber
\end{eqnarray}
The operator $D'$, which is applied to $j$ to generate the map, and the uncertainty dispersion $D$ are not identical any more. Neither of them can still
be expressed as $(M+\sum_ip_i^{-1}S_i^{-1})^{-1}$, due to the operator structure of the $S_i^{-1} D$ and $S_i^{-1}m\,m^\dagger$ terms. This was also found in \cite{2010arXiv1002.2928E}.

However, if we can assume that this operator processes any channel in the $i$-th band in a similar way, we can replace $S_i^{-1}D\,S_i^{-1}$ and $S_i^{-1}m\,m^\dagger S_i^{-1}$ by their channel averaged values $\Tr(D S_i^{-1})\, S_i^{-1}/\varrho_i$ and $\Tr(m\,m^\dagger S_i^{-1})\, S_i^{-1}/\varrho_i$, respectively. This permits to identify spectral coefficients of $D = (M+\sum_ip_i^{-1}S_i^{-1})^{-1}$ and $D' = (M+\sum_i{p'_i}^{-1}S_i^{-1})^{-1}$. For Jeffreys prior they become
\begin{eqnarray}
  p_i &=& \frac{\Tr(B_i)}{\varrho_i} \left[ 1 
-  \frac{2}{\varrho_i}\,\left(\frac{\Tr\left(m\,m^\dagger S_i^{-1} \right) }{\Tr(B_i )}\right)^2  \right]^{-1}
\!\!\!\!\!\!\!\! ,\; \mbox {and}\nonumber\\
  p_i' &=& \frac{\Tr(B_i)}{\varrho_i} \left[ 1 
+  \frac{2}{\varrho_i}\,\frac{\Tr\left(m\,m^\dagger S_i^{-1} \right)\, \Tr\left(D\, S_i^{-1} \right)}{\left(\Tr(B_i )\right)^2 } \right]^{-1},
\end{eqnarray}
where $m$, $D$, and  $B_i=(m\,m^\dagger+D) S_i^{-1}$ all depend on $p$. It is obvious, that the second order correction increases $p_i$ by some margin compared to \eq{eq:p_i0}, meaning that the reconstruction uncertainty increases. It is less obvious how $p'_i$ develops, since at first glance it seems to be corrected downwards. Note however, that an increased $p_i$ implies an increased $\Tr(B_i)$, since $D$ grows (spectrally) with increasing $p_i$.

The fact that we get two differing sets of spectral coefficients, $p_i$ and $p'_i$, reminds us to regard them as auxiliary variables of our signal reconstruction algorithm, rather than as optimal spectrum estimates.

\subsection{Poisson log-normal distribution with unknown spectrum}
\label{sec:LNPandspecrec}
The combined problem, reconstructing a Poisson log-normal signal with unknown spectrum, can now be treated approximatively. The combined free energy for the Gaussian posterior approximation to zeroth order is
\begin{eqnarray}
\tilde{G}(m,D) 
&\approx&\sum_i \left[ R_i^\dagger e^{b\,m + \frac{b^2}{2} \,\widehat{D}} - d_i\, \log\left( R_i^\dagger e^{b\,m}\right)\right] \nonumber\\
&+& \sum_i  \gamma_i\,   \log\left(q_i+ \frac{1}{2} \, \Tr\left((m\,m^\dagger +D)\, S_i^{-1} \right)\right) \nonumber\\
&-& \frac{T}{2} \Tr\left(1+\log(2\pi\,D)\right). 
\end{eqnarray}
The resulting map and uncertainty dispersion are provided by \eq{eq:pln0thorder} with the addition that $S=\sum_i\,p_i\,S_i$ and the $p_i$s are provided by \eq{eq:pi0thOrder}. Higher order corrections can be included in a similar way as in the individual problems. Also background counts with known or unknown covariance structure can be included in the same way they were treated in Sect. \ref{sec:obswithbackground}.

\section{Information synthesis}
\label{sec:infosynth}
\subsection{Multi-temperature posterior}
Although the obtained Gaussian knowledge states from minimal free energy estimation are approximative and therefore of limited accuracy, they might permit us to construct more accurate models of the posterior.
The idea is to combine several Gaussian distributions to a more accurate approximation of the true non-Gaussian posterior probability, and to measure the mean map and its uncertainty dispersion from this combination.

We recall that Gaussian approximations of the posterior obtained at low temperatures ($T\ll 1$) mostly carry information on its peak region, while those obtained at large temperatures ($T\gg 1$) information on its asymptotics. 
Also the canonical $T=1$ does not provide a perfect representation of the posterior, as a Gaussian approximation for a non-Gaussian PDF never can. 
However, by combining such different approximations in an appropriate way, we should obtain an improved representation of the correct PDF, which permits much easier calculation of moments like the signal mean and its uncertainty variance.

To this end we postulate the existence of a temperature distribution function $\p(T)$, such that
\begin{equation}
 \p(s|d) = \int_0^{\infty} \!\!\!\! dT \, \G(s-m_{(d,T)},D_{(d,T)})\, \p(T) 
\end{equation}
combines the different Gaussians with means $m_{(d,T)}$ and dispersions $D_{(d,T)}$ to synthesize the right posterior probability. A formal proof of the existence of $\p(T)$, and the necessary conditions for this is beyond the scope of this work. 
It should be noted, that e.g. multi-peaked distributions cannot accurately be represented by approximate Gaussians obtained at different temperatures. They can, however, often be well approximated by Gausians centered on those peaks. The recipes described below do not depend on the way the different Gaussians used in the mixture model were obtained, and therefore can also be used in such cases.

In the following we provide a recipe to construct $\p(T)$ in practice. We assume that $m_i = m_{(d,T_i)}$ and $D_i = D_{(d,T_i)}$ have been computed for a number $N_T$ of temperatures $T_i$. The temperatures are best chosen to sample well the different part of the posterior, its peak by having some $T_i \ll 1$, the bulk of the PDF with $T_i =1$, and the PDF tails with $T_i\gg 1$.

The surrogate probability function  we want to construct, and which should resemble the exact one as closely as possible, is therefore of the form
\begin{equation}
 \tilde{\p}(s|d) = \sum_{i=1}^{N_T} \G(s-m_i,D_i)\, P_i .
\end{equation}
$\tilde{\p}(s|d)$ should be as close as possible to $\p(s|d)$ in an information theoretical sense. The natural choice for the distance measure is the Kullback-Leibler divergence, which measures the cross-information of $\tilde{\p}(s|d)$ on $\p(s|d)$, and which is practically identical to the free energy $\tilde{G}[\tilde{\p}(s|d)] $ of our surrogate posterior according to \eq{eq:FreeeKL}.
Introducing un-normalized probabilities $p_i$ as our degrees of freedom, and setting $P_i = p_i/Z_p$ with $Z_p = \sum_j\, p_j$ in order to enforce the proper normalization, $\sum_i\,P_i =1$, this reads
\begin{equation}\label{eq:KLaction}
 \tilde{G}(p) = \sum_i \frac{p_i}{Z_p} (U_i - \tilde{U}_i(p)) - F.
\end{equation}
We have introduced the here irrelevant, since $p$-independent, free energy $F = - \log Z_d$ of the original problem and the energies $U_i$ and $ \tilde{U}_i(p)$ with respect to the template distributions $\G_i(s)= \G(s-m_i,D_i)$:
\begin{eqnarray}
 U_i &=& \expectWRT{ H(d,s) }{\G_i} = \int \!\!\mathcal{D}s \,\, \G_i(s)\, H(d,s) \;\mbox{and}\nonumber\\
 \tilde{U}_i(p) &=& \expectWRT{ \tilde{H}_p(s) }{\G_i},\;\mbox{with}\\
\tilde{H}_p(s) &=& - \log(\sum_i \, {p_i \, \G_i(s)}/Z_p).
\nonumber
\end{eqnarray}
\subsection{Minimizing the Gibbs energy}
\subsubsection{Analytical scheme}
Now one has to minimize $\tilde{G}(p)$ with respect to $p$. The problem to calculate the path integrals defining the energies was already addressed in this work. A systematic way is to Taylor-Fr\'{e}chet expand the Hamiltonians around the centers of the Gaussians $m_i$ and then use the known moments of $\G_i(s)$ to approximate the energies. For the surrogate energies this yields up to second order in $\phi_i = s - m_i$
 \begin{eqnarray}\label{eq:Uip}
 \tilde{U}_i(p) &=& - \log g_i + 
\frac{1}{2} \, 
\sum_j \frac{g_{j\, i}}{g_i}  \,\Tr(D_j^{-1} D_i)  \\
&+&
\frac{1}{2} \, 
\sum _{j\, k}\frac{g_{j\, i}}{g_i}  \,\left( \frac{g_{k\, i}}{g_i} - \delta_{jk}\right)  \,  m_{ij}^\dagger D_j^{-1} D_i \, D_k^{-1}  m_{ik}, 
\nonumber 
 \end{eqnarray} 
with
\begin{eqnarray}
g_{j\, i} &=& p_j \, \G_j(m_i)/Z_p, \;\mbox{and}\nonumber\\
 g_i &=& \sum_{j=1}^{N_j}\, g_{j\,i},\;\mbox{and}\\
 m_{ij} &=& m_i - m_j.\nonumber
 \end{eqnarray}
\subsubsection{Monte-Carlo scheme}
Alternatively,  one  can approximate the average $\expectWRT{ X[s] }{\G_i}$ of a quantity $X[s]$ by sums over $N_i$ sampling points $\{s_{i}^{(j)}\}_j$, which can easily be drawn from $\G_i(s)$:
\begin{equation}
 \expectWRT{ X[s] }{\G_i}\approx \sum_j \, X[s_{i}^{(j)}]/N_i .
\end{equation}
This way, $\tilde{G}(p)$ can be approximated, and minimized with a
suitable optimization scheme. 
The sampling points, their Gaussian probabilities $\G_{k\, i}^{(j)}= \G_k(s_{i}^{(j)})$,
 as well as the energies $U_i$ need only be calculated once, 
but the surrogate energies $U_i(p) = \log Z_p - \sum_j \, \log(\sum_k \, p_k\, \G_{k \, i}^{(j)})   /N_i$ 
have to be updated at any step of the scheme. 

One might argue, that if we use stochastic methods to build $\tilde{\p}(s|d)$, one could have used a Markov-Chain
Monte-Carlo (MCMC) method right from the beginning for the signal inference problem. However, we expect that the
here described posterior synthesis method should reproduce the correct posterior better than a sample point cloud,
since we are using well adapted Gaussians as our building blocks and not delta functions as the direct MCMC
approach uses. Furthermore, the analytical and sampling method can be combined, in that the analytical estimates
are combined with the sampling estimates of the contributions of the neglected terms in the Taylor-Fr\'echet
expansions of \eq{eq:Uip}. And finally, since our scheme draws samples from Gaussians, it can be trivially parallelized, which is not easily possible with MCMC schemes.

\subsection{Maps and moments}

Once the minimum of $\tilde{G}(p)$ with respect to $p$ is found, 
one has synthesized a posterior approximation with a Gaussian mixture model. 
From this, any moment of the distribution function can easily be calculated. 
The mean map can be expressed as
\begin{equation}
 \label{eq:fieldmeanformula}
m \approx \expectWRT{ s }{\tilde{P}(s)} = \sum_i \, P_i \expectWRT{ s }{\G_i(s)} = \sum_i \, P_i \, m_i,
\end{equation}
as well as the uncertainty dispersion as
\begin{equation}
 \label{eq:fielddispformula}
D \approx \expectWRT{ (s-m) \, (s-m)^\dagger }{\tilde{P}(s)} =  \sum_i \, P_i \,(D_i+ m_i\,m_i^\dagger) - m\,m^\dagger.
\end{equation}
We leave the verification and application of the information synthesis method for future work.

\section{Conclusions}
\label{sec:conclusions} 

We have shown that the minimal free Gibbs energy principle in information field theory can be used to obtain approximate knowledge states with maximal cross-information to the exact posterior. The construction of such knowledge states with Gaussian PDF is relatively straightforward:
\begin{enumerate}
 \item The joint PDF of signal and data $\p(d,s)$ has to be specified, e.g. by specifying a data likelihood $\p(d|s)$ and signal prior $\p(s)$, and using $\p(d,s)= \p(d|s)\,\p(s)$.
 \item The information Hamiltonian $H$ is the negative logarithm of this, $H(d,s)= -\log(\p(d,s))$.
 \item A suitably parametrized PDF as a surrogate for the posterior has to be specified, e.g. a Gaussian with its mean and dispersion as degrees of freedom.
 \item The internal energy $U$ and entropy $S_\mathrm{B}$ of this PDF have to be calculated as the PDF-average of the Hamiltonian and the negative log-PDF, respectively.
\item The Gibbs free energy, $G=U-T\,S_\mathrm{B}$, has then to be minimized with respect to all degrees of freedom of the surrogate PDF.
\item Any statistical summary like mean and variance can now be extracted from the surrogate PDF.
\end{enumerate}
The minimal free energy principle is therefore well suited to tackle statistical inference problems. We have demonstrated this with two different problems and their combination: reconstructing a log-normal field from Poisson data subject to a point spread function and reconstruction without prior knowledge on the signal power spectrum. Earlier results from renormalization calculations in \cite{2009PhRvD..80j5005E,2010arXiv1002.2928E} have been reproduced. The there used renormalization schemes can therefore be understood as aiming for a surrogate Gaussian PDF which has maximal cross information to the correct posterior. Since these results were previously shown to reconstruct well, also the here proposed method for the more complicated combined case  can be expected to work. However, a detailed implementation and verification of this was left for future work. 

Finally we have sketched how Gaussian knowledge states obtained at different thermodynamical temperatures can be combined into a more accurate representation of the posterior, from which moments of the signal uncertainty distributions can easily be extracted. 

The minimal Gibbs energy and maximal cross information principle introduced here to IFT should allow the construction of novel reconstruction schemes for statistical inference problems on spatially distributed signals. The thermodynamical language may help to clarify concepts and to simplify applications of IFT, since it permits us to tackle non-linear inverse problems without the need to use diagrammatic perturbation theory and renormalization schemes. 

\begin{acknowledgements}
We thank Mona Frommert, Jens Jasche, Niels Oppermann, Gerhard B{\"o}rner, and three referees for discussions and comments on the manuscript.
\end{acknowledgements}

\bibliography{../Bib/ift}
\bibliographystyle{apsrev}

\end{document}